\documentclass[prl,twocolumn,floats,aps,epsfig,nofootinbib,amssymb]{revtex4}
\usepackage{subfigure}
\usepackage{graphicx}
\usepackage{cancel}
\usepackage{amssymb}
\usepackage{textcomp}
\usepackage{amsmath}
\usepackage{bm}
\usepackage{times}
\usepackage{epsfig}
\usepackage{color}
\begin{document}
\title{\Large Three Layers of Neutrinos}
\author{Vernon Barger}
\author{Pavel Fileviez P{\'e}rez}
\author{Sogee Spinner}
\address{
Department of Physics, University of Wisconsin, Madison, WI 53706, USA}
\date{\today}
%%%%%%%%%%%%%%%%%%%%%%%%%%%%%%%%%%%%%%
\begin{abstract}
In this letter we point out that in a class of models for spontaneous R-parity breaking based on gauged B-L, 
the spectrum for neutrinos is quite peculiar. We find that those models generally predict three layers of neutrinos: 
one heavy sterile neutrino, two massive active neutrinos, and three nearly massless (one active and two sterile) 
neutrinos.  
\end{abstract}
\maketitle
%%%%%%%%%%%%
\textit{\bf Introduction}:
%%%%%%%%%%%%
The existence of massive neutrinos have motivated an infinite number of studies in physics 
beyond the standard model. Today, we know with good precision the values of the mass squared differences and mixing angles: 
$\Delta m_{12}^2=(7.58 \pm 0.21) \times 10^{-5} \rm{eV}^2$, 
$|\Delta m_{23}^2|=(2.40 \pm 0.15) \times 10^{-3} \rm{eV}^2$ for the solar and atmospheric
mass splittings, $\tan^2 \theta_{12} =0.484 \pm 0.048$, 
$\sin^2 2 \theta_{23}=1.02 \pm 0.04$ for the solar and atmospheric mixing angles, and 
$\sin^2 2\theta_{13}=0.07 \pm 0.04$~\cite{Strumia-Vissani}. 
Unfortunately, as in the case of the quark sector of the Standard Model (SM), 
we do not know the underlying theory for neutrino masses which could explain 
the above values and predict the type of spectrum: 
Normal Hierarchy (NH), Inverted Hierarchy (IH) or Quasi-Degenerated (QD). 

Recently, we have investigated in great detail the different theories where one can 
understand the origin of R-parity (non)conservation in supersymmetry~\cite{LR,PRL,Others,Fate}. 
In this context, we have found that if one sticks to the minimal model, R-parity must be 
broken spontaneously by the vacuum expectation values of the ``right-handed" sneutrinos 
present in the theory. These theories based on gauged B-L make a great number of predictions 
relevant for the discovery of supersymmetry at the Large Hadron Collider since the symmetry 
breaking scale (\textit{i.e.} the $Z^{'}$ mass), and the R-parity breaking scale are defined 
by the SUSY breaking scale which is at a TeV.  Proton decay~\cite{review} in these frameworks occur 
only through  higher-dimensional operators and the gravitino can be a good dark matter candidate. 
See Refs.~\cite{LR,PRL,Others} for details and for general aspects of R-parity violation in SUSY 
see Ref.~\cite{Barbier}. For an early study of spontaneous R-parity violation see Ref.~\cite{Murayama}.

In this letter we investigate the predictions for the neutrino spectrum in the class of 
models mentioned above. We point out that in these models for spontaneous R-parity breaking in supersymmetry, 
the spectrum for neutrinos is special. We find that these models generally predict three layers of neutrinos: 
one heavy sterile neutrino, two massive active neutrinos, and three nearly ``massless" (one active and two sterile) 
neutrinos. Therefore, the spectrum can be: $m_6 \gg m_{2,3} \gg m_1,m_4,m_5$ in the NH 
and $m_6 \gg m_{1,2} \gg m_3,m_4,m_5$ in the IH. Here, $m_6$ is the mass of the 
heavy neutrino with mass around GeV-TeV, $m_{1,2,3}$ are the masses of the active neutrinos, and $m_{4,5}$ 
correspond to the masses of the light sterile neutrinos.

%%%%%%%%%%%%%%%%%%%%%%%%%%%%%%%%%%%%%%%%%
\textit{\bf Theoretical Framework}:
%%%%%%%%%%%%%%%%%%%%%%%%%%%%%%%%%%%%%%%%%
It is well-known that the relevant scale for neutrino mass generation through the seesaw 
mechanism~\cite{seesaw} is typically defined by the $B-L$ scale. Here B and L stand for 
baryon and lepton number, respectively. The global $B-L$ in the Standard Model (SM) 
is an accidental anomaly free symmetry, and once it is gauged,  the $U(1)_{B-L}^3 \text{ and } U(1)_{B-L}$ 
anomalies must be cancelled. This is independent of  $B-L$ being embedded in a larger 
gauge group or combined with the weak hypercharge. The anomaly $B-L$ cancellation 
conditions can be satisfied by introducing three generations of right-handed neutrinos, 
which are singlets under the SM.

The Minimal Supersymmetric Standard Model (MSSM) particle content plus 
3 right-handed neutrinos have the following $B-L$ charges:
\begin{eqnarray}
	Q \sim \frac{1}{3}, \ & u^c \sim -\frac{1}{3}, \ &\  d^c \sim -\frac{1}{3},
	\notag
	\\
	L \sim -1,  \ & \  \nu^c \sim 1, \ & \ e^c \sim 1,
\end{eqnarray}
while the MSSM Higgses are neutral. Now, if we gauge $B-L$ in this context one can 
understand the origin of the (non)conservation of R-parity. This symmetry is defined as 
$R=(-1)^{2S + 3 (B-L)}$, where S is the spin of the particle. Since the crucial phenomenological 
and cosmological predictions of supersymmetric theories depend of the (non)conservation 
of this symmetry one should understand this crucial issue.

The breaking of local $B-L$ can be achieved either by introducing new vector-like pairs 
of Higgses (chiral superfields) with $B-L$ quantum numbers or by considering the minimal 
particle content~\cite{PRL}.  The latter case necessitates a vacuum expectation value (VEV) 
for the right-handed sneutrinos. Since these have an odd charge, $R$-parity will be spontaneously 
broken resulting in a theory that has some commonality with bilinear $R$-parity violating models, 
see Ref.~\cite{PRL}. Then, the non-trivial issue here is to give a large VEV to the right-handed sneutrinos.
The solution to this problem was proposed for the first time in Ref.~\cite{LR}, a tachyonic mass term is needed, 
and later in Ref.~\cite{Fate} we showed how to dynamically generate this mass through the radiative 
symmetry breaking mechanism.

In order to understand the predictions in the neutrino sector, one must study the symmetry
breaking and we proceed by examining the different contributions to the potential in the minimal
case.  In this context the superpotential is given by~\cite{PRL}
\begin{align}
	{\cal W} & = {\cal W}_{\text{MSSM}} \ + \ Y_\nu L H_u \nu^c,
	\notag
	\\
	{\cal W}_{\text{MSSM}} & = Y_u Q H_u u^c \ + \ Y_d Q H_d d^c \ + \ Y_e L H_d e^c \ + \  \mu H_u H_d,
\end{align}
where the only difference from the MSSM is the addition of a Yukawa coupling between 
the neutrinos and $H_u$. This gives us an $F$-term contribution to the potential which reads as
\begin{align}
	V_F & \supset |Y_\nu|^2 |\tilde \nu^c|^2 \left(|H_u^0|^2 + |\tilde \nu|^2\right)
		+ \left(-Y_\nu \mu^* \tilde \nu H_d^{0*} \tilde \nu^c + \rm{h.c.} \right).
\end{align}
The relevant contribution from the $D$-terms is
\begin{align}
	V_D \supset \frac{1}{8}g_{X}^2 \left(x_{\nu^c} |\tilde \nu^c_i|^2 + D_{\text{MSSM}}\right)^2.
\end{align}
Here, we have assumed that $B-L$ is a part of a group $U(1)_X$, where a particle $\psi$ has a 
charge $x_\psi$. $D_\text{MSSM}$ represents contributions from MSSM fields such as left-handed
sneutrinos or the Higgses which are neutral under $B-L$ but might still have a nonzero $x$-charge. 
Finally, the SUSY breaking potential contributes:
\begin{align}
V_\text{Soft} \supset \left(m_{\tilde \nu^c}^2\right)_{ij} \tilde \nu^c_i \tilde \nu^{c*}_j
			+ \left(a_\nu \tilde L H_u \tilde \nu^c + \rm{h.c.}\right) \ + \ \ldots,
\end{align}
where the soft mass matrix for the right-handed sneutrinos can be diagonalized without
loss of generality.
We must now consider which are the most important contributions to the scalar potential. 
The Yukawa coupling of the neutrinos to $H_u$ will lead to a neutrino Dirac mass term.  
Since the only scale in this potential is the SUSY breaking mass scale, any potential seesaw 
will have a maximum scale of a few TeV. Therefore, $Y_\nu$ must still be rather small for 
neutrino masses, $Y_\nu \lesssim 10^{-6}-10^{-5}$, which can be neglected when 
minimizing for $\tilde{\nu}^c$. The VEV of the left-handed sneutrino also contributes 
to neutrino masses through a seesaw mediated by the gauginos, placing an upper 
bound around, $\langle \tilde{\nu} \rangle \lesssim 10^{-3}$ GeV, these terms can be ignored 
as well.

The trilinear $a_\nu$ term helps to determine $\langle \tilde{\nu} \rangle$ 
and so must also be negligible. This leaves the relevant part of the potential as
\begin{align}
	V & = \frac{1}{8}g_X^2 \left(x_{\nu^c} |\tilde \nu^c_i|^2 + D_{\text{MSSM}}\right)^2
		+ \left(m_{\tilde \nu^c}^2\right)_{i} |\tilde \nu^c_i|^2.
\end{align}
Specifically, what we would like to know is how many of the right-handed sneutrinos
attain a VEV.  Minimizing with respect to $\tilde \nu^c_j$ yields
\begin{equation}
	\frac{1}{4} g_X^2 x_{\nu^c} \tilde \nu^{c*}_j \left(x_{\nu^c} |\tilde \nu^c_i|^2 + D_{\text{MSSM}}\right)
	\ + \  \left(m_{\tilde \nu^c}^2\right)_{j} \tilde \nu^{c*}_j = 0,
\end{equation}
where there is no sum over $j$. At least one non-trivial solution exists if the soft mass squared is negative.  
Since we have three equations for the same quantity, two options exist: 
assume that all the soft masses are equal and the three equations become one 
or return to the case where only one right-handed sneutrino VEV is non-zero. 
Notice that in the former case, the potential has an $U(3)_{\tilde{\nu}^c}$ 
flavor symmetry, and one can always rotate to a basis where only one of the 
right-handed sneutrinos acquires a VEV.  In general, the other two generations 
will also get VEVs but these will be smaller than even the VEVs of the left-handed sneutrinos. 
The upshot of this discussion is that lepton number is only broken in one family and therefore 
only one of the right-handed neutrinos can get a TeV scale mass leaving the other two 
masses to be determined by the parameters responsible for the active neutrino masses. 
Therefore: \textit{the minimal SUSY $B-L$ models or the simplest theories for R-Parity 
predict the existence of 2 sterile neutrinos which are 
degenerate or lighter than the active neutrinos.} 

%%%%%%%%%%%%%%%%%%%%%%%%%%%%%%%%%%%%%%%%%
\textit{\bf Neutrino Mass Spectrum}:
%%%%%%%%%%%%%%%%%%%%%%%%%%%%%%%%%%%%%%%%%
We Continue in the $B-L$ scenario ~\cite{PRL} for simplicity.
In the basis: $(\nu_i, \nu^c_I, \tilde B', \tilde B, \tilde W, \tilde H_d^0, \tilde H_u^0)$, the neutralino mass matrix is
\begin{widetext}
\begin{equation}
\label{N.Mass}
\mathcal{M} = 
\begin{pmatrix}
	0_{3 \times 3}
	&
	\frac{(Y_\nu)_{jI} \, v_u}{\sqrt 2}
	&
	-\frac{g_{BL} v_{Lj}}{2}
	&
	-\frac{g_1 v_{Lj}}{2}
	&
	\frac{g_2 v_{Lj}}{2}
	&
	0_{3 \times 1}
	&
	\frac{(Y_\nu)_{jK} \, v_{RK}}{\sqrt 2}
\\
	\frac{(Y_\nu)_{Ji}  v_u}{\sqrt 2}
	&
	0_{3 \times 3}
	&
	\frac{g_{BL} v_{RJ}}{2}
	&
	0_{3 \times 1}
	&
	0_{3 \times 1}
	&
	0_{3 \times 1}
	&
	\frac{(Y_\nu)_{kJ} v_{Lk}}{\sqrt 2}
\\
	-\frac{g_{BL}  v_{Li}}{2}
	&
	\frac{g_{BL}  v_{RI}}{2}
	&
	M_{BL}
	&
	0
	&
	0
	&
	0
	&
	0
\\
	-\frac{g_{1} v_{Li}}{2}
	&
	0_{1 \times 3}
	&
	0
	&
	M_1
	&
	0
	&
	-\frac{g_1 v_d}{2}
	&
	\frac{g_1 v_u}{2}
\\
	\frac{g_2 v_{Li}}{2}
	&
	0_{1 \times 3}
	&
	0
	&
	0
	&
	M_2
	&
	\frac{g_2 v_d}{2}
	&
	-\frac{g_2 v_u}{2}
\\
	0
	&
	0
	&
	0
	&
	-\frac{g_1 v_d}{2}
	&
	\frac{g_2 v_d}{2}
	&
	0
	&
	-\mu
\\
	\frac{(Y_\nu)_{Ki} \, v_{RK}}{\sqrt 2}
	&
	\frac{(Y_\nu)_{Ik} \, v_{Lk}}{\sqrt 2}
	&
	0
	&
	\frac{g_1 v_u}{2}
	&
	-\frac{g_2 v_u}{2}
	&
	-\mu
	&
	0
\end{pmatrix}.
\end{equation}
\end{widetext}
Here,
\begin{eqnarray}
v_R &=&\frac{\sqrt{-8 m_{\tilde{\nu}^c}^2}}{ g_{BL}}, 
\\
v_L &=&\frac{B_\nu \ v_R}{ \left( m_{\tilde L}^2 - \frac{1}{8} g_{BL}^2 v_R^2 \right)},
\end{eqnarray}
with $B_\nu = \frac{1}{\sqrt{2}} \left( Y_\nu \mu v_d - a_\nu v_u \right)$.
Again, $m_{\tilde{\nu}^c}^2 < 0$, see Refs.~\cite{LR,Fate}. As discussed above, only one 
generation of the ``right-handed" sneutrinos gets a TeV size VEV, while the rest are 
quite small and so we proceed with $v_{R1} = v_{R2} = 0$ and $v_{R} \equiv v_{R3} \neq 0$. 
For an early study of the generation of neutrino masses using the VEV of right-handed sneutrinos 
see Refs.~\cite{Murayama,Mohapatra}.
One can neglect the active neutrinos to try to understand how many heavy states exist in
$\mathcal{M}$.  We know that this number is equal to five (for the neutralinos) plus
some number of the right-handed neutrinos.  The characteristic polynomial reveals that
only one right-handed neutrino attains a large mass, hence verifying the earlier claim that only one generation gets a large Majorana mass. 
The mass matrix $\mathcal{M}$ then contains six heavy states, 
which can be integrated out using the seesaw mechanism to approximate the masses of the
light neutrinos (three active and two sterile):
\begin{equation}
\label{seesaw}
\mathcal{M}_\nu = m - m_D M^{-1} m_D^T,
\end{equation}
where the light Majorana mass matrix is given by
\begin{equation}
m = 
\begin{pmatrix}
	0_{3 \times 3}
	&
	\frac{(Y_\nu)_{i\beta} v_u}{\sqrt 2}
\\
	\frac{(Y_\nu)_{\alpha i} v_u}{\sqrt 2}
	&
	0_{2 \times 2}
\end{pmatrix}.
\end{equation}
Here, $\alpha$ and $\beta$ run over the two light right-handed neutrinos. 
The heavy Majorana mass matrix is given by
\begin{widetext}
\begin{equation}
M = 
\begin{pmatrix}
	0
	&
	\frac{g_{BL} v_R}{2}
	&
	0
	&
	0
	&
	0
	&
	\frac{(Y_\nu)_{k3} \,  v_{Lk}}{\sqrt 2}
\\
	\frac{g_{BL} v_R}{2}
	&
	M_{BL}
	&
	0
	&
	0
	&
	0
	&
	0
\\
	0
	&
	0
	&
	M_1
	&
	0
	&
	-\frac{g_1 v_d}{2}
	&
	\frac{g_1 v_u}{2}
\\
	0
	&
	0
	&
	0
	&
	M_2
	&
	\frac{g_2 v_d}{2}
	&
	-\frac{g_2 v_u}{2}
\\
	0
	&
	0
	&
	-\frac{g_1 v_d}{2}
	&
	\frac{g_2 v_d}{2}
	&
	0
	&
	-\mu
\\
	\frac{(Y_\nu)_{3k} \, v_{Lk}}{\sqrt 2}
	&
	0
	&
	\frac{g_1 v_u}{2}
	&
	-\frac{g_2 v_u}{2}
	&
	-\mu
	&
	0
\end{pmatrix},
\end{equation}
and finally the Dirac mass matrix is
\begin{equation}
m_D =
\begin{pmatrix}
	\frac{(Y_\nu)_{i3}  v_u}{\sqrt 2}
	&
	-\frac{g_{BL}  v_{Li}}{2}
	&
	-\frac{g_1  v_{Li}}{2}
	&
	\frac{g_2  v_{Li}}{2}
	&
	0_{3 \times 1}
	&
	\frac{(Y_\nu)_{i3}  \, v_{R}}{\sqrt 2}
\\
	0_{2 \times 1}
	&
	0_{2 \times 1}
	&
	0_{2 \times 1}
	&
	0_{2 \times 1}
	&
	0_{2 \times 1}
	&
	\frac{(Y_\nu)_{k\alpha} \, v_{Lk}}{\sqrt 2}
\end{pmatrix}.
\end{equation}
\end{widetext}
Before studying the specific form of the resulting five-by-five mass matrix, it is useful to
examine the matrix in a very general way.  For this purpose it can be cast as
\begin{equation}
	\label{genform}
	\mathcal{M}_\nu =
	\begin{pmatrix}
		M^{LL}
		&
		M^{LR}
	\\
		{M^{LR}}^T
		&
		M^{RR}
	\end{pmatrix},
\end{equation}
where $M^{LL}$ is the three-by-three left-handed Majorana mass matrix, $M^{RR}$ 
is the two-by-two right-handed Majorana mass matrix and $M^{LR}$ is the Dirac mixing 
between the left-handed and right-handed sectors. Such a form allows for four possibilities:
\begin{enumerate}
	\item
	{
	\label{Mixed}
		Mixed: $M^{LR} \sim M^{LL} \ \text{or} \ M^{RR}$.
	}
	\item
	{
	\label{Dirac}
		Pseudo-Dirac: $M^{LR} \gg M^{LL}, M^{RR}$.
	}
	\item
	{
	\label{MajoranaL}
		``Inverted" seesaw: $M^{LL} \gg M^{LR}, \ M^{RR}$.
	}
	\item
	{
	\label{MajoranaR}
		``Traditional" seesaw $M^{RR} \gg M^{LR}, \ M^{LL}$.
	}
\end{enumerate}
Case \ref{Mixed} is already ruled out since it would allow for large mixings between active and light sterile
neutrinos, which has not been experimentally observed. Case \ref{Dirac} leads to one Majorana 
left-handed neutrino and two so called pseudo-Dirac neutrinos. The pseudo-Dirac neutrinos have a 
small mass splitting between the active and sterile neutrinos on the order of magnitude of the 
Majorana mass and with a near maximal mixing.  Because of this, the mass splitting is 
severally restricted, see for example Ref.~\cite{deGouvea:2009fp}. Cases \ref{MajoranaL} and 
\ref{MajoranaR} allows for all active and sterile neutrinos to be Majorana. Here active-sterile mixings 
will be controlled by the Dirac mass and therefore this quantity will be bounded by data on 
neutrino mixing.  We will see that only Case \ref{MajoranaL} is possible in our models.

The specific forms of these submatrices are linear combinations of matrices that are made up of 
products of elements with flavor structure, \textit{i.e.} Yukawa couplings and sneutrino VEVs,
where the coefficients are ratios of gaugino/Higgsino masses.  $\mathcal{M}_\nu$ can be 
treated as an expansion in the flavorful parameters, where to leading order we keep terms with a maximum product of two
flavorful parameters.  Then, one gets the following mass matrix for the light (active and sterile) neutrinos:
\begin{widetext}
\begin{equation}
\label{Mnu}
M_\nu =
\begin{pmatrix}
	A \ v_{Li} v_{Lj}
	+ B \ \left[(Y_\nu)_{i3} v_{Lj} + (Y_\nu)_{j3} v_{Li} \right]
	+ C \ (Y_\nu)_{i3} (Y_\nu)_{j3} 
	&
	\frac{1}{\sqrt{2}} v_u (Y_\nu)_{i \beta} 
\\
	\frac{1}{\sqrt{2}} v_u (Y_\nu)_{\alpha j}
	&
	D \left[(Y_\nu)_{k \alpha} \ v_{Lk}\right] \left[(Y_\nu)_{m \beta} \ v_{Lm}\right]
\end{pmatrix},
\end{equation}
where
\begin{align}
	A & = \frac{2 \mu^2}{\tilde m^3}, \ \ \
	B = \left(\frac{v_u}{\sqrt{2} v_R}  + \frac{\sqrt{2} \mu v_d v_R}{\tilde m^3}\right),
	\ \ \
	C = \left(\frac{2 M_{BL} v_u^2}{g_{BL}^2 v_R^2} + \frac{v_d^2 v_R^2}{\tilde m^3}\right), \ \ \
	D  =\frac{v_d^2}{\tilde m^3},
\end{align}
\begin{align}	
	\tilde m^{3} & = \frac
		{
			4
			\left[
				\mu v_u v_d \left(g_1^2 M_2 + g_2^2 M_1 \right)
				- 2 M_1 M_2 \mu^2
			\right]
		}
		{g_1^2 M_2 + g_2^2 M_1}
		.
\end{align}
\end{widetext}
From the experimental upper limits on active neutrino masses we obtain $(Y_\nu)_{i \alpha} \lesssim 10^{-12}$.
This can be compared to $(Y_\nu)_{i 3} \lesssim 10^{-5}$, which is less constrained because of the TeV scale
seesaw suppression.  
The active neutrino masses must come from $M^{LL}$, and it is worthwhile to diagonalize that 
submatrix first. This can be done by rotating $M_\nu$ by
\begin{equation}
	V_1 = 
	\begin{pmatrix}
		U_{ij}
		&
		0_{2\times3}
		\\
		0_{3\times2}
		&
		1_{2\times2}
	\end{pmatrix},
\end{equation}
where $U$ 
is a three-by-three submatrix that diagonalizes $M^{LL}$.  It is easy to show that
because of the flavor structure of $M^{LL}$, \textit{i.e.} only two flavor dependent parameters: 
$v_{Li}$ and $(Y_\nu)_{i3}$, it has a zero eigenvalue.  Then, only the  
NH and IH  are allowed. Assuming $M^{RR}= 0$,
since it is fourth order in the small parameters (indicating case 4 is not possible), yields
\begin{equation}
	V_1^T  M_\nu V_1  = 
	\begin{pmatrix}
		M^{LL}_D
		&
		\frac{1}{\sqrt{2}} v_u U^T Y_\nu
		\\
		\frac{1}{\sqrt{2}} v_u Y_\nu^T U
		&
		0_{2\times2}
	\end{pmatrix}, 
\end{equation}
where $M^{LL}_{D} = \text{Diagonal}(0,m_2,m_3) \text{ and } \text{Diagonal}(m_1,m_2,0)$ for NH and IH respectively.  Case 2 is not possible here since $M_{ii}^{LL}$ will would lead to large pseudo-Dirac splitting, then we are left with case 3 as stated earlier.  The eigenvalues of the above are two massive active neutrinos and three neutrinos (one active and two steriles) 
with mass around or bellow the active neutrino scale. This proves our main statement: 
\textit{minimal SUSY $B-L$ models or the simplest theories for R-Parity imply the existence 
of 2 sterile neutrinos which are degenerate or lighter than the active neutrinos.} 
The second case is the most natural scenario.

%%%%%%%%%%%
\textit{\bf Summary}
%%%%%%%%%%
In this letter we have pointed out that in a class of models for spontaneous R-parity breaking in supersymmetry, 
the spectrum for neutrinos is quite peculiar. We find that these models generally predict three layers of neutrinos: 
one heavy sterile neutrino, two massive active neutrinos, and three near massless (one active and two sterile) 
neutrinos. One can have: $m_6 \gg m_{2,3} \gg m_1,m_4,m_5$ in the NH scenario and 
$m_6 \gg m_{1,2} \gg m_3,m_4,m_5$ in the IH case. Here $m_6$ is the mass of the heavy neutrino 
with mass around GeV--TeV, $m_{1,2,3}$ are the masses of the active neutrinos, and $m_{4,5}$ are the 
masses of the light sterile neutrinos. We will discuss the implications for colliders and long baseline neutrino experiments elsewhere.

%%%%%%%%%%%%%%%%%%%%%%%%%%%%%%%%%%%%%%%%%%%%%%%%%%%%%%%%%%%%%%%%%%%%%%%%
{\textit{Acknowledgments}}:
{\small The work of V.B., P.F.P., and S.S. is supported in part by 
the U.S. Department of Energy under grant No. DE-FG02-95ER40896,
and by the Wisconsin Alumni Research Foundation.}
Note: Ref.~\cite{Goran}, which appeared on the same day as this paper, also studies this topic.

%%%%%%%%%%%%%%%%%%%%%%%%%%%%%%%%%%%%%%%%%%%%%%%%%%%%%%%%%%%%%%%%%%%%%%%%%


\begin{thebibliography}{000}
%%%%%%%%%%%%%%%%%%%%%%%%%%%%%%%%%%
%\cite{Strumia:2006db}
\bibitem{Strumia-Vissani}
  A.~Strumia and F.~Vissani,
  ``Neutrino masses and mixings and..,''
  arXiv:hep-ph/0606054.
  %%CITATION = HEP-PH/0606054;%%

\bibitem{LR}
  P.~Fileviez P\'erez and S.~Spinner,
  ``Spontaneous R-Parity Breaking and Left-Right Symmetry,''
  Phys.\ Lett.\  B {\bf 673} (2009) 251;
  [arXiv:0811.3424 [hep-ph]].
  %%CITATION = PHLTA,B673,251;%%
 %\cite{Barger:2008wn}
  
 \bibitem{PRL}
  V.~Barger, P.~Fileviez Perez and S.~Spinner,
  ``Minimal gauged $U(1)_{B-L}$ model with spontaneous R-parity violation,''
  Phys.\ Rev.\ Lett.\  {\bf 102}, 181802 (2009);
  [arXiv:0812.3661 [hep-ph]].
  %%CITATION = PRLTA,102,181802;%% %\cite{FileviezPerez:2009gr}
 
 \bibitem{Others}
  P.~Fileviez Perez and S.~Spinner,
  ``Spontaneous R-Parity Breaking in SUSY Models,''
  Phys.\ Rev.\  D {\bf 80}, 015004 (2009);
 % [arXiv:0904.2213 [hep-ph]].
  %%CITATION = PHRVA,D80,015004;%%
  %\cite{Everett:2009vy}
  %\bibitem{Everett:2009vy}
  L.~L.~Everett, P.~Fileviez Perez and S.~Spinner,
  ``The Right Side of Tev Scale Spontaneous R-Parity Violation,''
  Phys.\ Rev.\  D {\bf 80}, 055007 (2009);
 % [arXiv:0906.4095 [hep-ph]].
  %%CITATION = PHRVA,D80,055007;%%
  %\cite{FileviezPerez:2009bw}
  %\bibitem{FileviezPerez:2009bw}
  P.~Fileviez Perez and S.~Spinner,
  ``TeV Scale Spontaneous R-Parity Violation,''
  arXiv:0909.1841 [hep-ph].
  %%CITATION = ARXIV:0909.1841;%%

%\cite{FileviezPerez:2010ek}
\bibitem{Fate}
  P.~Fileviez Perez and S.~Spinner,
  ``The Fate of R-Parity,''
  arXiv:1005.4930 [hep-ph].
  %%CITATION = ARXIV:1005.4930;%%
 
%\cite{Nath:2006ut}
\bibitem{review}
  P.~Nath and P.~Fileviez P\'erez,
  ``Proton stability in grand unified theories, in strings, and in branes,''
  Phys.\ Rept.\  {\bf 441} (2007) 191;
% [arXiv:hep-ph/0601023].
  %%CITATION = PRPLC,441,191;%%
  %\cite{FileviezPerez:2004th}
%\bibitem{FileviezPerez:2004th}
  P.~Fileviez P\'erez,
  ``How large could the R-parity violating couplings be?,''
  J.\ Phys.\ G {\bf 31} (2005) 1025.
 % [arXiv:hep-ph/0412347].
  %%CITATION = JPHGB,G31,1025;%%

\bibitem{Barbier}
  R.~Barbier {\it et al.},
  ``R-parity violating supersymmetry,''
  Phys.\ Rept.\  {\bf 420} (2005) 1.
%  [arXiv:hep-ph/0406039].
  %%CITATION = PRPLC,420,1;%%
  
 %\cite{Hayashi:1984rd}
\bibitem{Murayama}
  M.~J.~Hayashi and A.~Murayama,
  ``Radiative Breaking Of $SU(2)_R \times U(1)_{B-L}$ Gauge Symmetry Induced By Broken
  $N=1$ Supergravity In A Left-Right Symmetric Model,''
  Phys.\ Lett.\  B {\bf 153} (1985) 251.
  %%CITATION = PHLTA,B153,251;%%

  \bibitem{seesaw}
  P.~Minkowski,
  ``Mu $\to$ E Gamma At A Rate Of One Out Of 1-Billion Muon Decays?,''
  Phys.\ Lett.\ B {\bf 67} (1977) 421;
  %%CITATION = PHLTA,B67,421;%%
  T. Yanagida,
in {\it Proceedings of the Workshop on the Unified Theory
   and the Baryon Number in the Universe}, eds. O. Sawada et al.,
p.~95, KEK Report 79-18 (1979);
  M. Gell-Mann, P. Ramond and R. Slansky,
   in {\it Supergravity}, eds. P. van Nieuwenhuizen et al.,
   (North-Holland, 1979), p.~315;
  R.~N.~Mohapatra and G.~Senjanovi\'c,
  ``Neutrino Mass And Spontaneous Parity Nonconservation,''
  Phys.\ Rev.\ Lett.\  {\bf 44} (1980) 912.
  %%CITATION = PRLTA,44,912;%
  
  %\cite{Mohapatra:1986aw}
\bibitem{Mohapatra}
  R.~N.~Mohapatra,
  ``Mechanism for understanding small neutrino mass in superstring theories,''
  Phys.\ Rev.\ Lett.\  {\bf 56} (1986) 561.
  %%CITATION = PRLTA,56,561;%%
  
%\cite{deGouvea:2009fp}
\bibitem{deGouvea:2009fp}
  A.~de Gouvea, W.~C.~Huang and J.~Jenkins,
  ``Pseudo-Dirac Neutrinos in the New Standard Model,''
  Phys.\ Rev.\  D {\bf 80}, 073007 (2009)
%  [arXiv:0906.1611 [hep-ph]].
  %%CITATION = PHRVA,D80,073007;%%
  
%\cite{Ghosh:2010hy}
\bibitem{Goran}
  D.~K.~Ghosh, G.~Senjanovi\'c and Y.~Zhang,
  ``Naturally Light Sterile Neutrinos from Theory of R-parity,''
  arXiv:1010.3968 [hep-ph].


\end{thebibliography}
\end{document}